# Self-Directed Channel Memristor for High Temperature Operation

Kristy A. Campbell

*Dept. of Electrical and Computer Engineering*
*Boise State University*
*1910 University Dr.*
*Boise, ID 83725*
email: kriscampbell@boisestate.edu

*Abstract*—Ion-conducting memristors comprised of the layered chalcogenide materials $Ge_2Se_3/SnSe/Ag$ are described. The memristor, termed a self-directed channel (SDC) device, can be classified as a generic memristor and can tolerate continuous high temperature operation (at least 150 °C). Unlike other chalcogenide-based ion conducting device types, the SDC does not require complicated fabrication steps, such as photodoping or thermal annealing, making these devices faster and more reliable to fabricate. Device pulsed response shows fast state switching in the $10^{-9}$ s range. Device cycling at both room temperature and 140 °C show cycling lifetimes of at least 1 billion.

## I. INTRODUCTION

Memristors [1] have been studied intensely for the past several years due to their potential use in applications such as non-volatile memory [2], neuromorphic and bio-inspired computing [3-8], and threshold logic [9].

The type of memristor described in this work is an ion-conducting device which relies on $Ag^+$ movement into channels within the device active layer to change the device resistance. This memristor, referred to as a self-directed channel (SDC) device, uses a metal-catalyzed reaction within the device active layer to generate permanent conductive channels that contain Ag agglomeration sites. The amount of Ag within the channel determines the resistance of the device.

In this work, electrical properties of the layered memristor device are presented. These include the response of the device to a quasi-static DC IV sweep as a function of temperature and compliance current, frequency response to a sinusoidal input signal, lifetime cycling, and pulsed response.

## II. DEVICE OPERATION

Before describing the operation of the SDC memristor, it must be noted that this device should not be confused with another type of ion-conducting device which also uses Ag or Cu ions to change device resistance, referred to as the 'conductive bridge' device (often referred to as CBRAM and also as programmable metallization cell, PMC). The CBRAM device changes resistance through a mechanism involving the formation and dissolution of a conductive filament between the top and bottom electrodes in response to a potential applied across the device [10]. Similarities between the SDC and CBRAM devices are that both typically use chalcogenide materials as the active layer such as $As_xS_y$ [11], AgInSbTe [12], $Ge_xSe_y$, or $Ge_xS_y$ [13, 14] and both use an easily oxidizable metal, such as Cu or Ag, to change the device conductivity [2, 10, 11-17]. However, in the CBRAM device a Cu or Ag metal layer in contact with the chalcogenide layer is the source of Cu or Ag metal ions generated by an applied potential across the device. These ions migrate toward the more negative electrode under an applied potential, where they get reduced and build-up a metallic filament towards the positive electrode which eventually bridges the two electrodes and reduces the device resistance [10]. Reversing voltage polarities between the electrodes causes the conductive filament to disperse, thus increasing the device resistance.

Another difference between the SDC and CBRAM type of chalcogenide-based ion-conducting devices is that the CBRAM is typically fabricated using Se-rich or S-rich glasses by either depositing a ternary material (e.g. Ge-S-Ag) to a desired stoichiometry [16], or by photodoping and/or thermally annealing the Ag or Cu metal into the active amorphous material matrix [9, 10, 13,16, 21-23]. To achieve the proper concentration of metal in the glass, precise control of the amount of metal included in the chalcogenide and the stoichiometry of the chalcogenide material is required. Both of these are challenging to achieve and are critical to the consistent operation of the CBRAM device [18-20]. In addition, the photodoping/annealing fabrication methods significantly reduce the maximum temperature exposure of the device during operation and fabrication. Two major factors that contribute to this are: 1) once the Ag or Cu has been added to the material, reduction in the glass transition temperature occurs and with exposure to higher temperatures this can result in crystallization of the glass, which damages device functionality; and 2) the chalcogenides are prone to over saturation by diffusion of the metal layer into the active layer at higher temperatures.

The SDC device, Fig. 1, uses a Ge-rich chalcogenide glass, $Ge_2Se_3$, and no photodoping or thermal annealing. The device is operational immediately after fabrication. The $Ge_2Se_3$ active layer is where device switching occurs; the key feature of this material is the presence of Ge-Ge homopolar bonds. The three layers consisting of $Ge_2Se_3/Ag/Ge_2Se_3$, directly below the top W electrode, mix together during deposition and jointly form the Ag-source layer. This Ag-source layer is not in direct contact with the active layer. This allows the device to have significantly higher processing and operating temperatures (above 250 °C and at least 150 °C, respectively) since Ag does not migrate into the active layer at high temperatures, and the active layer maintains a high glass transition temperature (~350 °C). These processing and operating temperatures are higher than most ion-conducting chalcogenide device types, including the S-based glasses (*e.g.* GeS) that need to be photodoped or thermally annealed. It is a combination of these factors that allow the SDC device to operate over a wide range of temperatures, including long-term continuous operation at 150 °C. The SnSe layer assists in the selective incorporation of Ag ions into the $Ge_2Se_3$ layer during operation by incorporation of Sn ions during the first forming step of the device near the regions of Ge-Ge bonding within the $Ge_2Se_3$ layer [26 – 28].

It should be noted that while the layered SDC device structure looks complicated due to the number of material layers, it is actually simpler and more reliable to fabricate than the CBRAM device. The entire deposition of the film layers, including the top electrode, is done *in-situ* in one processing step using a standard sputter tool. No extra time is required for photodoping or annealing, as is needed for the CBRAM device. Additionally, layer thicknesses are not critical; the active layer could be considered the only thickness sensitive layer, but it has a wide margin of acceptable variation, between 300 and 500 Å. Wafer-to-wafer consistency is therefore high, as across-wafer film thickness variation is not a factor. Because tight controls do not need to be in place for maintaining film thicknesses to tight tolerances, tool qualifications can be done less frequently and a production line could continue for longer periods without being out of specification. Because of this, wafer yields for the SDC devices can be relied on to be >90%.

In contrast, the CBRAM device depends critically on the amount of Ag incorporated into the device during photodoping/annealing. This means that the glass thickness and the Ag thickness need to be well-controlled as slight variations can cause the device to switch poorly (too little Ag) or to become saturated (too much Ag). The photodoping step also depends on good thickness control since the time of light exposure is linked to the amount of Ag incorporated into the device during processing. Consequently, the processes need to be frequently monitored and tools qualified more often. This translates to production down time and frequent poor yields. Additional complications of the CBRAM device fabrication include the device sensitivity to light exposure. As Ag can be photodoped into the device during light exposure, the wafers must be maintained in a dark environment until the risk of photodoping is removed.

In summary, five main factors differentiate the SDC device in terms of operation and fabrication:

1. The device can operate continuously at 150 °C without degradation.
2. No photodoping or thermal annealing is required, saving time, money, and handling risks.
3. The *in-situ* deposition of all device layers, including the top electrode, occurs with a standard sputter deposition tool in a single step.
4. Film thicknesses in the stack are not critical.
5. Cost is reduced due to a decrease in processing time, use of a single sputter deposition, reduction in qual/down time, and increased yield.

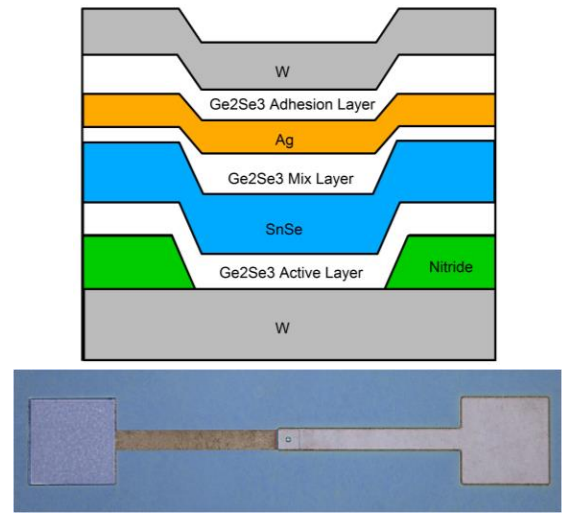

Fig. 1. Memristor device structure. Top: device layers. Bottom: fabricated device showing top electrode (right) and bottom electrode (left) bond pads for electrical probing. The layer thicknesses are not to scale relative to each other.

SDC devices are initially in a high resistance state (MΩ to GΩ range) following fabrication. The first time a device is operated after fabrication the device self-directed channel is formed during application of a positive potential to the top electrode. The potential required for this operation is the same as required during normal device operation. This first operation generates Sn ions from the SnSe layer and forces them into the active $Ge_2Se_3$ layer [24-26]. Theoretical calculations predict that these Sn ions facilitate the incorporation of Ag into the active layer at the Ge-Ge bonding sites [27]. This occurs through an energetically favorable process in which the electrons entering the active layer from the negative bottom electrode, concurrently with the formation of Sn ions near the positive top electrode, enable formation of a pair of self-trapped electrons in the $Ge_2Se_3$ active layer strongly localized around the Ge-Ge dimers present in this Ge-rich glass [28]. The result of this is that Sn ions facilitate an energetically favorable reaction of Ag substitution for Ge on the Ge-Ge bond. During this reaction, the glass network is distorted, creating an 'opening' near the Ge-Ge sites. The open regions provide good access for $Ag^+$ to the Ag-Ge site and become natural

'conductive channels' within the active layer for the movement of Ag$^+$ during device operation. This self-directed channel is a result of the natural glass structure and follows the location of the initial Ge-Ge dimers within the glass. Since Ag has a tendency to agglomerate with other Ag atoms, these sites encourage Ag agglomeration within the glass. Thus, device resistance changes by adding or removing Ag from the agglomeration sites within this in-situ generated pathway. It is expected then that conduction occurs between clusters of Ag agglomeration sites [29, 30]. This pathway does not therefore have to consist of conductive metallic filaments [31] spanning the two electrodes, as in the CBRAM device. It is simply a channel that has varying concentrations of Ag within it at these Ag agglomeration sites. The concentration of Ag at a given agglomeration site, and the distance between agglomeration sites dictates the device resistance. The resistance is tunable in the lower and higher directions by movement of Ag onto or away from these agglomeration sites through application of either a positive or negative potential, respectively, across the device.

## III. EXPERIMENTAL

### A. Device Structure and Fabrication

Ion-conducting devices were fabricated with a via structure and top and bottom electrodes, each of which extends to a metal pad for wirebonding or electrical probing access, Fig. 1, bottom. Devices were fabricated on 100 mm p-type Si wafers, in a stacked layer structure with a via defining the device contact size. Via sizes ranged from 0.25 μm to 4 μm in diameter. Device operation was independent of via size within this range.

Prior to deposition of the device material layers, the wafers were pre-sputtered with Ar$^+$ to remove any oxide species from the bottom electrode followed by *in-situ* sputter deposition of all of the remaining device layers and top W electrode layer, using an AJA International ATC Orion 5 UHV Magnetron sputtering system. The target layer thicknesses were (from bottom to top): Ge$_2$Se$_3$ (300 Å)/SnSe (800 Å)/Ge$_2$Se$_3$ (150 Å)/Ag (500 Å)/Ge$_2$Se$_3$ (100 Å)/W (400 Å). Final device etching was performed with a Veeco ME1001 ion-mill. The active switching layer is the 300 Å Ge$_2$Se$_3$ layer deposited adjacent to the bottom electrode.

### B. Electrical Measurements

Electrical measurements consisted of: DC measurements as a function of compliance current and temperature; continuous-wave (CW) response for memristor classification and for cycling measurements; and pulse response to single and consecutive programming pulses.

Electrical measurements were performed at the wafer level using a Micromanipulator 6200 microprobe station equipped with an MC-Systems temperature controllable wafer hot chuck (23, 50, 100, and 150 °C). Wafers were equilibrated for at least 30 minutes at the measurement temperature prior to all measurements. Device sizes tested were 250 nm diameter.

DC quasi-static sweep measurements were performed with an HP4156A semiconductor parameter analyzer. All DC sweep measurements consisted of a Write/Erase/Write sequence, where Write corresponds to a 0 to 1 V voltage sweep, and Erase corresponds to a 0 to −1 V voltage sweep.

Sinusoidal CW and pulsed measurements were made using an Agilent B1500A Semiconductor Parameter Analyzer equipped with two 2-channel Waveform Generator/Fast Measurement Units (WGFMUs). In the CW measurements, used to demonstrate the device classification of memristor type [1], a sinusoidal input signal was applied to the device for 10 cycles at each frequency starting from 0.5 Hz and increasing to 100 kHz. (Note: this sinusoidal input frequency response is not representative of the device speed or pulsed frequency response, but is instead a fingerprint classification of memristor type [1].) The WGFMUs allowed direct measurement of the current through the device during testing without external circuits or current limiting series resistors.

In the cycling test, an arbitrary waveform was created for an HP33250A arbitrary waveform generator, and used with an Agilent 54815A oscilloscope and a test circuit consisting of the memristor and a series load resistor. The load resistor ($R_{load}$ = 10 kΩ) was used to limit the current through the device during cycling measurements to prevent device damage. The input waveform frequency was 1 kHz. The oscilloscope was set to persistence mode during each test so that any variations in the device response could be observed. The device response was sampled from the oscilloscope every decade.

## IV. RESULTS AND DISCUSSION

### A. Quasi-Static DC Measurements

A typical quasi-static DC IV measurement curve is shown in Fig. 2. The device displays a bipolar IV curve, where the write sweep initiates the formation of a conductive channel through the device, resulting in a low resistance state, and the erase sweep results in a high resistance state via movement of Ag out of the channel. In this device type, at normal operating temperatures up to at least 150 °C, the first DC write sweep applied to a device post fabrication, usually called the forming sweep [10], does not require application of a higher potential like other memristor device types.

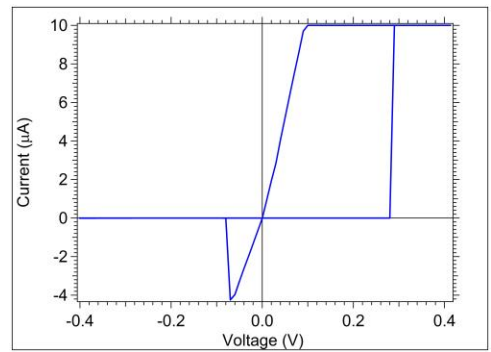

Fig. 2. Typical quasi-static DC IV curve for the memristor device. The compliance current used in this measurement was 10 μA.

The value of the compliance current during the write sweep has a significant effect on the programmed low resistance state (Fig. 3) [17, 21, 30]. However, the erased state is only significantly impacted after a write sweep that uses a 1 mA compliance current. In this case, the erased resistance does not typically exceed 200 kΩ, and the impact on those devices is permanent.

Devices maintained at a temperature of 150 °C for two days and then returned to room temperature show no difference in DC IV response compared to devices that have not previously been heated. The programmed resistance for this case as a function of compliance current is also shown in Fig. 3.

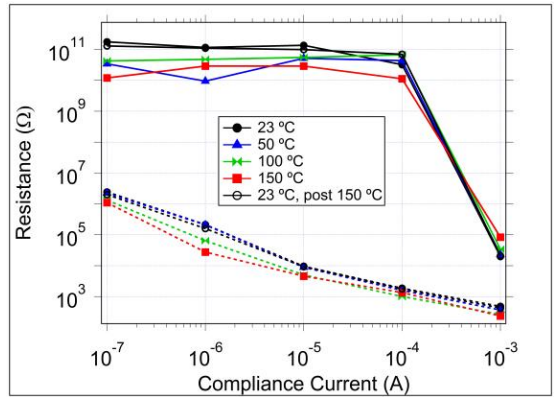

Fig. 3. Resistance as a function of compliance current and operating temperature. Dashed lines represent the written resistance states. Solid lines represent the erased resistance states following a write at the given compliance current.

### B. Device Response to Sinusoidal Input

The IV curves showing the device response to a sinusoidal input at frequencies ranging from 0.5 Hz to 100 kHz are shown in Fig. 4. All curves pass through the origin for every frequency. As the frequency is increased, the IV curve lobe area decreases, eventually collapsing into a line. This is the characteristic of a generic memristor [1]. The response of the device to this sinusoidal input should not be confused with the device response to programming pulses. This sinusoidal input test is used to classify the memristor type.

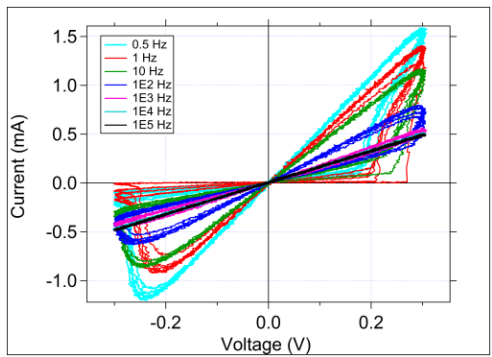

Fig. 4. IV curve generated with a sinusoidal input with frequency varying from 0.5 Hz to 100 kHz. Ten cycles at each frequency are shown.

### C. Lifetime Cycling

Devices were cycled at room temperature and 140 °C, using an arbitrary waveform (Fig. 5 (a) and (b), respectively). The input waveform is shown as the solid black trace for both cases. The input waveform for the 140 °C test includes a low amplitude read pulse after each write and erase peak in order to verify the programmed state during operation. The other traces in each graph, one trace collected after each decade of cycling, correspond to the voltage drop across the load resistor. The voltage drop across the memristor is the difference between the input signal and the signal across the load resistor. When the resistance of the memristor is low, the voltage drop across the load resistor is high; when memristor resistance is high, the voltage drop across the load resistor is low. The traces at each temperature show that both devices cycle up to at least one billion cycles.

The erase response clearly shows conduction up to an erase threshold voltage (the small peak in the load resistor voltage at the start of the erase input peak) after which the memristor resistance increases and the voltage across the load decreases.

The read pulse during the 140 °C measurement is used to verify the state of the device after programming. In all cases, the read pulse shows the device was programmed as expected.

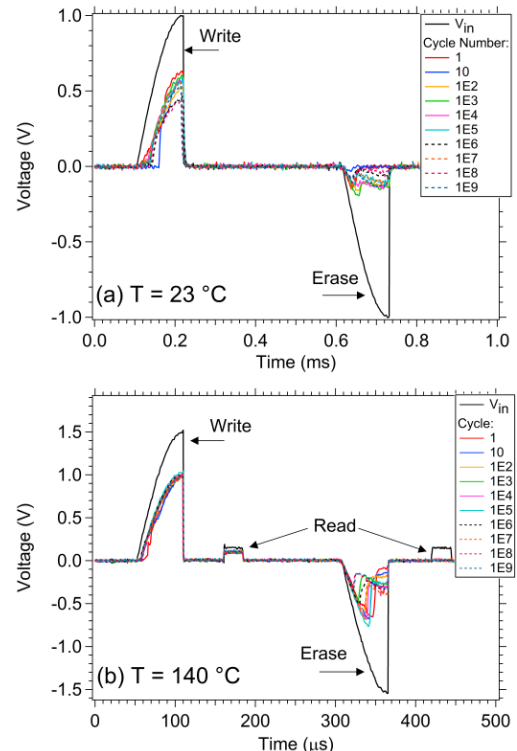

Fig. 5. Continuous cycling. A sample of the device response was saved at each decade of the measurement. (a) 23 °C and (b) 140 °C.

### D. Pulsed Response

Device response to a sequence of alternating 70 ns full-width-half-max (FWHM) single erase and write pulses is shown in Fig. 6. In this example, the write and erase pulse amplitudes were selected to be large enough to write the device into a very low resistance state and to erase it to a very high resistance state for binary operation. For this test, the device was initially programmed to a low resistance state which was then measured prior to application of the first 'erase' programming pulse using a 'read' pulse. The read pulse, at a low enough potential not to perturb the resistance state of the device, measures the current directly through the device. The first read pulse in Fig. 6 shows approximately 200 µA, which corresponds to a low resistance value of 1 kΩ. Following the initial read pulse, a negative voltage 'erase' pulse is applied to place the device into a high resistance state. The read pulse following the erase pulse measures current within the noise floor, indicating that the device has fully erased to a high resistance state. In this case, the current is at or below the minimum detectable signal, which is > 1 MΩ for the measurement range used during this test.

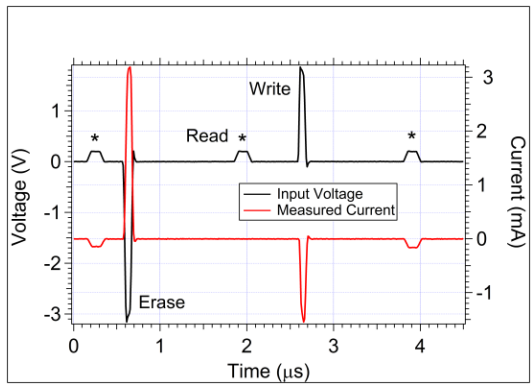

Fig. 6. Erase/Write pulse response for binary operation of the memristor.

The large amplitude positive 'write' pulse drives the device back into a low resistance state (Fig. 6). The read following this write pulse indicates that the device was again written to approximately 1 kΩ. Repeated application of the Erase/Write binary pulse sequence cycles the device between high and low resistance states.

Even though the programming amplitudes used in Fig. 6 were selected to provide binary operation, this device is also capable of being programmed through a continuous range of resistances in both the high and low resistance programming directions. To achieve continuous resistance operation, the write and erase pulse amplitudes are reduced from the case of binary operation and consecutive blocks of write and erase pulses are applied. The pulse width can also be varied to assist in achieving this response. Both of these are demonstrated in Fig. 7, where a change in the erase pulse amplitude from −3 V to −1 V, the write amplitude from 1 V to 0.5 V, and the pulse widths from 70 to 500 ns FWHM, results in continuous resistance programming. The consecutive pulsing sequence shown uses a series of eight read/erase pulse pairs followed by eight read/write pulse pairs. In the sequence in Fig. 7, the first read shows an initial device resistance of 12 kΩ. Each subsequent erase pulse increases the resistance incrementally. After the eighth erase pulse in Fig. 7, the resistance has increased to approximately 60 kΩ. Similarly, the following sequence of write pulses reduces the resistance incrementally from 60 kΩ to 9 kΩ.

## V. CONCLUSIONS

Electrical characteristics of SDC $Ge_2Se_3$/SnSe/Ag-based ion-conducting memristor devices were described. The SDC devices offers two major advantages over other types of ion-conducting chalcogenide devices: 1) Continuous operation at 150 °C; and 2) no photodoping required during fabrication. The device fabrication process is simple, with one *in-situ* film deposition step using a standard commercially available sputter tool for all device layers, including the top electrode.

The DC programmed resistance as a function of compliance current and operating temperature, up to 150 °C, shows no significant variation in programmed resistance as a function of temperature. Furthermore, device cycling at room temperature and 140 °C both show functional devices out to at least 1 billion cycles.

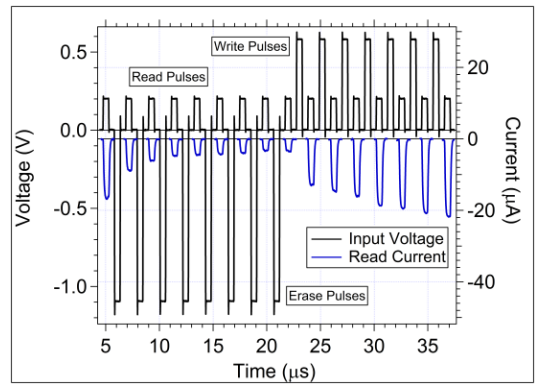

Fig. 7. Continuous resistance programming using eight consecutive erase pulses followed by eight consecutive write pulses. Each programming pulse is followed by a 200 mV read pulse. Only the current measured during the read pulse is shown for clarity.

These devices are classified as generic memristors based on their response to a sinusoidal input signal over a frequency range of 0.5 Hz to 100 kHz.

The devices can be programmed over a continuous range of resistance states using two techniques: DC compliance current limiting and pulsed operation. Consecutive pulsing can selectively place a device into a desired resistance range, either through consecutive erase or write pulses. This range can be selected by varying the number of pulses applied, the pulse width, and/or the pulse amplitude. The data retention in each intermediate resistance state is currently under study and depends upon the conditions used to program the device.

## ACKNOWLEDGMENTS

This work was partially supported by a grant from the National Science Foundation, grant no. CCF-1320987, the United States Air Force Office of Scientific Research, DEPSCoR Grant No. FA9550-07-1-0546, and by the United

States Air Force Research Laboratory, Grant No. FA9453-08-2-0252.